%% file: main.tex
\documentclass[aip,reprint]{revtex4-1}

\usepackage[utf8]{inputenc} % allow utf-8 input
\usepackage[T1]{fontenc}    % use 8-bit T1 fonts
\usepackage{hyperref}       % hyperlinks
\usepackage{url}            % simple URL typesetting
\usepackage{booktabs}       % professional-quality tables
\usepackage{amsfonts}       % blackboard math symbols
\usepackage{nicefrac}       % compact symbols for 1/2, etc.
\usepackage{microtype}      % microtypography
\usepackage{graphicx}
\usepackage{lipsum}
\usepackage{comment}
\usepackage[caption=false]{subfig}
\usepackage{textcomp}
\usepackage{soul}           % highlight text \hl{...} and strike \st{...}
\usepackage{amsmath}

\usepackage[dvipsnames]{xcolor}

\usepackage{chngcntr}
\usepackage{enumitem}
\usepackage[table,xcdraw]{xcolor}
\usepackage{booktabs}
\usepackage{multirow}
\usepackage{hhline}
\usepackage{amssymb}
\usepackage{siunitx}

\usepackage{bibunits}
\defaultbibliography{references}      % your .bib file
\defaultbibliographystyle{apsrev4-1}      % or the journal-required style

\usepackage{euler}

\definecolor{codegreen}{rgb}{0,0.45,0}
\definecolor{codegray}{rgb}{0.5,0.5,0.5}
\definecolor{backcolour}{rgb}{0.95,0.95,0.95}

\usepackage{listings}
\usepackage{inconsolata}

\lstdefinestyle{mystyle}{
    backgroundcolor=\color{backcolour},
    commentstyle=\color{codegreen},
    keywordstyle=\color{blue},
    numberstyle=\tiny\color{codegray},
    basicstyle=\ttfamily\footnotesize,
    breakatwhitespace=false,
    breaklines=true,
    captionpos=b,
    keepspaces=true,
    numbers=left,
    numbersep=6pt,
    showspaces=false,
    showstringspaces=false,
    showtabs=false,
    tabsize=2,
    frame=none,
    xleftmargin=0pt,
    xrightmargin=0pt
}
\let\oldtextbf=\textbf
\renewcommand*{\textbf}[1]{\ifmmode\mathbf{#1}\else\oldtextbf{#1}\fi}
\renewcommand*{\phi}[0]{\varphi}

\newcommand{\paragraphtitle}[1]{\textsf{\textbf{\small {#1}}}}

\draft % marks overfull lines with a black rule on the right

\begin{document}
% Use the \preprint command to place your local institutional report number 
% on the title page in preprint mode.
% Multiple \preprint commands are allowed.
%\preprint{}

\title{Let's Stalk About Membranes: Committor-Based Enhanced Sampling of Stalk Formation} %Title of paper

% repeat the \author .. \affiliation  etc. as needed
% \email, \thanks, \homepage, \altaffiliation all apply to the current author.
% Explanatory text should go in the []'s, 
% actual e-mail address or url should go in the {}'s for \email and \homepage.
% Please use the appropriate macro for the type of information

% \affiliation command applies to all authors since the last \affiliation command. 
% The \affiliation command should follow the other information.

\author{Giorgia Rossi}
%\homepage[]{Your web page}
%\thanks{}
%\altaffiliation{}
\affiliation{Atomistic Simulations, Italian Institute of Technology, 16156 Genova, Italy}
\affiliation{Department of Physics, University of Genova, 16146 Genova, Italy}
% \altaffiliation{These authors contributed equally}

\author{Enrico Trizio}
%\homepage[]{Your web page}
%\thanks{}
%\altaffiliation{}
\affiliation{Atomistic Simulations, Italian Institute of Technology, 16156 Genova, Italy}
% \altaffiliation{These authors contributed equally}

\author{Davide Bochicchio}
%\homepage[]{Your web page}
%\thanks{}
%\altaffiliation{}
% \email[]{caius@iit.it}
\affiliation{Department of Physics, University of Genova, 16146 Genova, Italy}
\affiliation{INFN sezione di Genova, 16146 Genova, Italy}
% \altaffiliation{These authors contributed equally}
% \affiliation{These authors contributed equally}

\author{Giulia Rossi}
%\homepage[]{Your web page}
%\thanks{}
%\altaffiliation{}
% \email[]{caius@iit.it}
\affiliation{Department of Physics, University of Genova, 16146 Genova, Italy}
\affiliation{INFN sezione di Genova, 16146 Genova, Italy}
% \altaffiliation{These authors contributed equally}
% \affiliation{These authors contributed equally}

\author{Michele Parrinello$^*$}
\email[]{michele.parrinello@iit.it}
\affiliation{Atomistic Simulations, Italian Institute of Technology, 16156 Genova, Italy}

% Collaboration name, if desired (requires use of superscriptaddress option in \documentclass). 
% \noaffiliation is required (may also be used with the \author command).
%\collaboration{}
%\noaffiliation

\date{\today}

\begin{abstract}
\noindent\textbf{Abstract}
\input{manuscript/abstract}
\end{abstract}

%\pacs{}% insert suggested PACS numbers in braces on next line

\maketitle %\maketitle must follow title, authors, abstract and \pacs

% % for splitting bib
% \begin{bibunit}

\input{manuscript/paper}

\section*{Code and Data Availability} \label{sec:code_avail}
    The committor models have been trained using the open-source library \texttt{mlcolvar}.~\cite{bonati2023mlcolvar}
    Enhanced sampling simulations have been performed using a custom interface for the PLUMED~\cite{plumed2019promoting} plugin for enhanced sampling and free energy calculations.
    Simulation inputs and the code needed for training and performing enhanced sampling simulations are available in the GitHub repository~\url{https://github.com/giorgiarossii/membrane_fusion_committor.git}.

\begin{acknowledgments}
        The authors are grateful to Luigi Bonati, Ioannis Galdadas and Umberto Raucci for useful feedback and discussions on the manuscript. 
        Giulia Rossi acknowledges funding from the Italian MUR, grant P2022EKHKL.
\end{acknowledgments}

\section*{Competing interests statement}
    The authors declare no competing interests.
%\clearpage
% Create the reference section using BibTeX:
\section*{References}
\bibliography{references}

% \putbib

% \bibliographystyle{plain}
% \end{bibunit}

\clearpage
\onecolumngrid
% \begin{bibunit}

% begin supporting info
% reset counters for supporting info
\setcounter{page}{1}
\renewcommand{\thepage}{S\arabic{page}}
\setcounter{section}{0}
\renewcommand{\thesection}{S\arabic{section}}
\setcounter{equation}{0}
\renewcommand{\theequation}{S\arabic{equation}}
\setcounter{figure}{0}
\renewcommand{\thefigure}{S\arabic{figure}}
\setcounter{table}{0}
\renewcommand{\thetable}{S\arabic{table}}
    
\clearpage
% \onecolumngrid

{\Large\normalfont\sffamily\bfseries{{Supporting Information}}}

% Set table variables
\setlength{\tabcolsep}{18pt}
\renewcommand{\arraystretch}{1.2}
% Set table variables

% start supporting info
\input{manuscript/supporting}
% end supporting info

\clearpage
% \section*{Supporting References}
% \bibliography{references}
% \putbib
\newpage

% \end{bibunit}

\end{document}

%% file: manuscript/abstract.tex
Membrane fusion is essential for cellular communication and function, and understanding how two lipid bilayers merge is key to informing therapeutic strategies. 
Functionalized nanoparticles have recently emerged as synthetic fusogens, but the molecular mechanisms driving this process remain unclear, partly because fusion involves transitions over high free-energy barriers, difficult to capture in molecular simulations.
While enhanced sampling methods can address this problem, they also rely on the definition of collective variables, which are especially hard to define for fusion, as it arises from the collective rearrangement of many molecules and cannot be easily reduced to a simple intuitive coordinate.
Here, we study stalk formation, the first step of fusion, mediated by an amphiphilic gold nanoparticle, by employing an enhanced sampling strategy based on the committor function, machine-learned through a self-consistent procedure.
This method requires minimal prior knowledge of the system and leverages the learned committor function as an effective collective variable, enabling uniform sampling of the entire pathway.
From the resulting reactive trajectories and extensive transition region sampling, we obtain converged free-energy estimates and mechanistic insight into stalk formation.

%% file: manuscript/paper.tex
%\section{Introduction}

Membrane fusion is the topological transformation by which two lipid bilayers merge, enabling lipid mixing and, ultimately, exchange of aqueous contents between the two original membrane-bound compartments. Although fusion is central to many cellular processes, from vesicular trafficking to membrane repair and viral entry, it is increasingly important also as an engineering principle. Synthetic fusogenic agents, including peptides, nucleic acids, polymers, lipid nanoparticles, and inorganic nanoparticles, are now being developed to trigger fusion in controlled settings for applications in synthetic biology, nanomedicine, and intracellular delivery\cite{marsden2011model,leonardini2026}. In this context, the central question is no longer simply how biological fusion occurs, but how molecular design parameters can be tuned to make fusion efficient, selective, and predictable.

Synthetic fusogens exploit different physical routes to favor membrane fusion. They may promote membrane adhesion and dehydration, generate curvature, perturb local lipid packing, e.g. by stabilizing lipid protrusions. Gold nanoparticles functionalized with mixed hydrophobic and charged ligands provide one representative example: their amphiphilic character enables membrane insertion and interactions with lipid headgroups, making them effective promoters of membrane fusion\cite{atukorale2018structure,tahir2020calcium, brosio2023nanoparticle,canepa2023cholesterol,leonardini2025physical}. However, more generally, synthetic fusogens span a broad design space, where size, shape, surface chemistry, flexibility, and membrane-anchoring geometry may all affect the fusion pathway. Rational design therefore requires a mechanistic understanding of how these molecular features reshape the fusion pathway and alter the stability of its intermediate states.

At the molecular level, fusion between lipid bilayers proceeds through collective rearrangements. Once apposed membranes are brought into close contact ($\sim$2 nm), local dehydration and lipid-tail protrusions can trigger the formation of the first hydrophobic bridge between bilayers. This is referred to as the stalk, a key kinetic bottleneck that may evolve towards the fusion pore \cite{morandi2022extracellular}. 
Molecular dynamics simulations can resolve this process at the nanoscale and, in presence of synthetic fusogens, connect fusogen design features to microscopic mechanisms. 
However, these transitions occur on timescales that exceed those of standard approaches, as they require crossing large free energy barriers. As a consequence, they are \textit{rare events} in simulations, thus making their direct observation challenging.
Conveniently, several strategies can be devised to alleviate this limitation by making simulations both faster and more efficient. 
For instance, significant speed-ups can be obtained using a coarse-grained (CG) representation of the system to describe the interatomic interactions that drive the dynamics. 
By reducing the number of degrees of freedom and smoothing the underlying energy landscape, CG models substantially extend the accessible length and timescales compared to all-atom force fields.
However, even with the use of CG models, the observation of spontaneous fusion events remains challenging in simulations. 

This calls for the use of enhanced sampling techniques~\cite{henin2022enhanced,zhu2026enhanced}, which add an external bias potential to the original energy landscape to lower effective barriers in order to observe reactive events, reconstruct free-energy landscapes, and characterize the transition state ensemble.
In many of these methods, the bias is expressed as a function of a few collective variables (CVs), which are continuous and differentiable functions of the atomic coordinates. 
Therefore, the effectiveness of these methods strongly relies on the definition of an appropriate CV. If the CV is poorly chosen, it may lead to hysteresis, overlook important slow degrees of freedom, and artificially collapse metastable and transition-state configurations into overlapping regions of the projected space.
Several enhanced sampling strategies have been applied to membrane fusion, including umbrella sampling\cite{dagostino2017tethering,risselada2014expansion,hub2017probinga,hub2021joint}, and path-finding approaches\cite{fuhrmans2015mechanics,muller2012transition,smirnova2019thermodynamically}. These methods, using CVs such as intermembrane distance, lipid-tail connectivity, local density fields, or path coordinates in a reduced CV space, have made it possible to compute free-energy profiles for the formation of early fusion intermediates and to clarify how lipid composition, hydration, and membrane stress affect the fusion barrier\cite{poojari2021free}. 

In this work, we adopt a different strategy based on the committor function $q(\textbf{x})$, which, given two metastable states $A$ and $B$, returns the probability that a trajectory initiated from configuration $\textbf{x}$ reaches state $B$ before passing by $A$~\cite{weinan2010transition}. As a consequence, it provides a rigorous, probabilistic description of reaction progress and is considered to be the optimal reaction coordinate. 
However, determining the committor is far from trivial. Recently, we have proposed a self-consistent iterative procedure in which machine learning and enhanced sampling are combined to leverage the variational principle the committor obeys~\cite{kang2024computing,trizio2025everything}. 
Besides learning a parametrization for the committor, this method allows the rare event to be characterized in detail, starting from information limited to the initial and final state of the process.
To this end, the committor is represented as a neural network $q(\textbf{d}(\textbf{x}))$ that takes as input a set of physically motivated descriptors $d(\textbf{x})$, and is optimized via a loss function encoding the variational principle. 
This, in practice, amounts to minimizing the expectation value of the gradients of the committor with respect to the atomic positions $\textbf{x}$, while enforcing appropriate boundary conditions.
A progressively improved estimate of the committor is obtained through an iterative, self-consistent procedure in which we alternate between sampling and learning.
At each iteration, the current estimate of the committor is used both as a CV to drive enhanced sampling simulations and to construct a bias potential aimed at increasing sampling of the otherwise unfavored transition state region.
The configurations thus sampled are then used as the training dataset for the next iteration step, and the cycle is iterated until convergence.

The original formulation of this approach has been successfully applied to a wide range of systems~\cite{kang2024computing,trizio2025everything,das2025machine,deng2026fluctuations,berselli2026}, but when the relevant degrees of freedom involve a large number of atoms, as is the case for membrane fusion, it can become too computationally demanding due to the explicit calculation of the gradients with respect to the atomic coordinates. 
To overcome this limitation, we recently introduced~\cite{trizio2026ceci} an approximated variational functional that bypasses the expensive calculation of coordinate gradients, and simply uses the much cheaper gradients with respect to the input descriptors. 
Although this approximated principle does not formally target the \textit{exact} committor function, it still retains the ability to enhance sampling and to characterize the TS ensemble, thus providing an efficient tool for the study of complex processes such as membrane fusion, where both the dimensionality of the system and the complexity of the relevant degrees of freedom pose significant challenges.

In this work, we apply this committor-based enhanced sampling framework to the study of stalk formation induced by a functionalized amphiphilic gold nanoparticle. The input descriptors we chose are indicators of the presence and quality of a hydrophobic connection between the bilayers, and are general enough to be transferable to the study of fusion induced by other synthetic or biogenic fusogens. We demonstrate that this approach provides an effective and semi-automatic tool to study stalk formation, able to characterize the complex dynamics of the process by uniformly sampling the whole reactive pathway, observing many reactive events and extensively sampling the transition state region.
As a result, we obtain accurate free energy profiles and perform an in-depth probability-based analysis of the different stages leading to stalk formation.
Being semi-automatic, fast, and easy to set up, this method provides a flexible, general tool for studying fusion processes.
\begin{figure}[h!]
    \centering
    \includegraphics[width=1\linewidth]{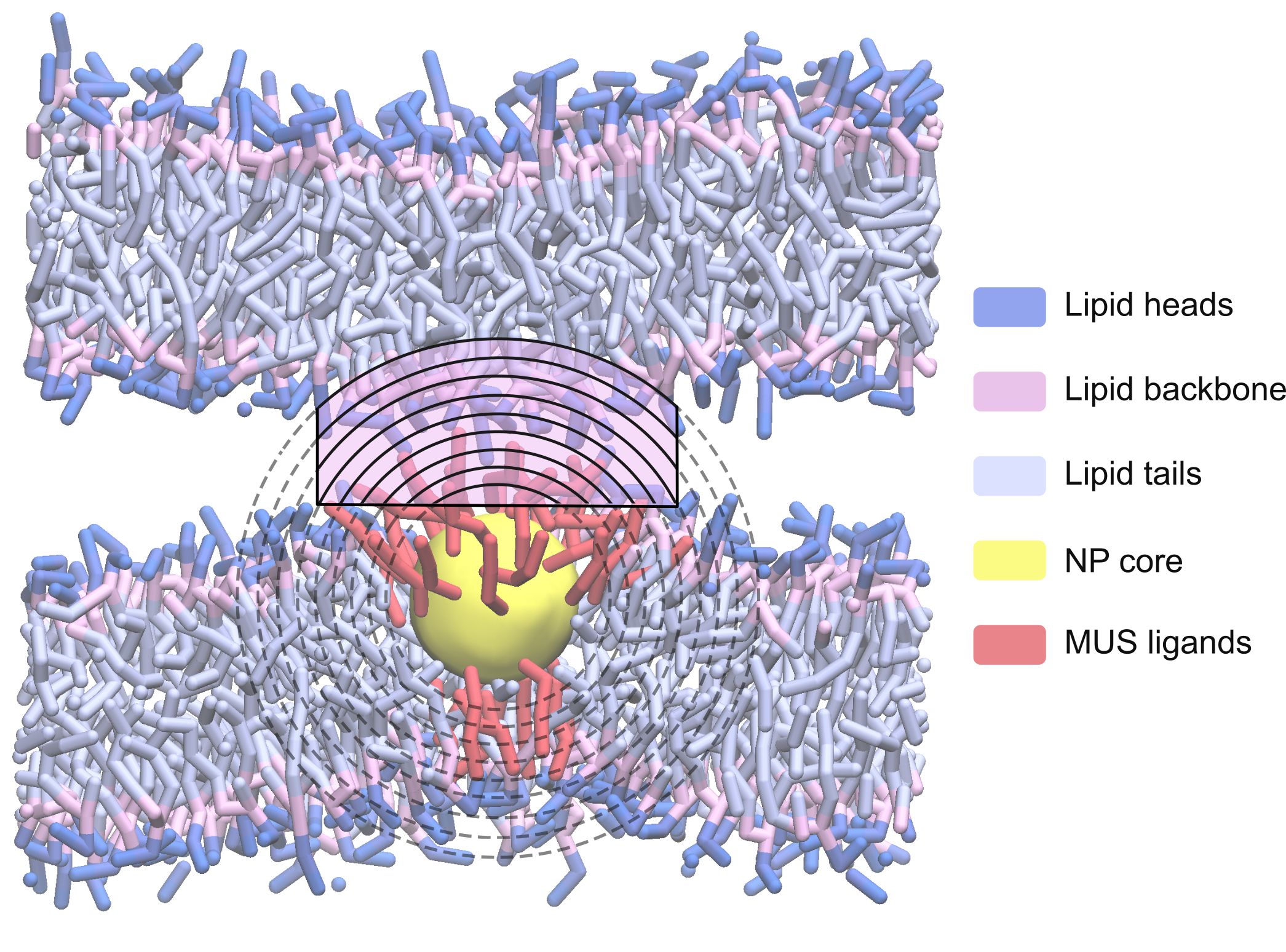}
    \caption{\textbf{Descriptor construction.} Cross-section of two apposed DOPC bilayers with a NP embedded in the lower membrane. Lipid tail coordination numbers are computed within concentric spherical shells centered on the NP center of mass. Only the shaded region is retained in the descriptor definition: a height cutoff excludes lipid tails below the NP surface, while a lateral cutoff confines the analysis to the intermembrane region directly above the NP, where stalk formation occurs.}
    \label{fig:descriptors}
\end{figure}

% \section{Methods}

% In the following, we summarize the committor-based enhanced sampling framework introduced in Refs.~\citenum{kang2024computing,trizio2025everything,trizio2026ceci} and employed here. 
% We first define the committor function and its machine learning representation. 
% We then describe the enhanced sampling scheme and, finally, summarize the self-consistent iterative procedure used to determine the committor. 

% \subsection{Committor function}
% \label{sec:committor}

\paragraphtitle{Method: Committor function.} 
Given two metastable states $A$ and $B$, the committor function $q(\mathbf{x})$ is defined as the probability that a trajectory initiated from configuration $\mathbf{x}$ reaches state $B$ before visiting $A$.
By construction, it satisfies the boundary conditions $q(\mathbf{x}_{i\in A})=0$ and $q(\mathbf{x}_{i\in B})=1$, and varies smoothly between these values across the transition state (TS) region. 
Owing to its probabilistic nature, the committor provides a rigorous description of the progress of a rare event and is widely regarded as the optimal reaction coordinate. 
However, its direct determination is challenging. 
One possible route to obtain it is to use a variational principle which, under the assumption of overdamped Langevin dynamics, states that the committor can be obtained by imposing the boundary conditions above and minimizing the functional $ K[q(\mathbf{x})] = \left\langle \left| \nabla_{\mathbf{u}} q(\mathbf{x}) \right|^2\right\rangle$, 
where $\nabla_{\mathbf{u}}$ denotes the gradient with respect to mass-weighted coordinates and $\langle \cdot \rangle$ is the Boltzmann average. 

% \subsection{Machine learning the committor}
% \label{sec:ml}
\paragraphtitle{Method: Machine learning the committor.}
Following Ref.~\citenum{kang2024computing}, we introduce a machine learning parametrization of the committor, which is progressively refined through alternating cycles of training and sampling.
In this framework, the committor is represented as a feed-forward neural network (NN) $q_{\boldsymbol{\theta}}(\mathbf{x}) = f_{\boldsymbol{\theta}}(d(\mathbf{x}))$, parametrized by a set of learnable weights $\boldsymbol{\theta}$ and taking as input a set of physical descriptors $d(\mathbf{x})$. 
The parameters $\boldsymbol{\theta}$ are optimized by minimizing a loss function $L = L_v + \alpha L_b$, composed of two terms scaled to have the same order of magnitude by the hyperparameter $\alpha$.
The \textit{boundary loss} $L_b$ enforces the correct boundary conditions (i.e., $q(\mathbf{x}_{i\in A})=0$ and $q(\mathbf{x}_{i\in B})=1$) and is evaluated on a labeled dataset of configurations collected from short unbiased simulations in the metastable basins and labeled accordingly.
The \textit{variational loss} $L_v$ encodes the variational principle for the committor discussed above and is evaluated on configurations with the associated statistical weights.
In its original formulation, $L_v$ requires computing the gradients of the committor with respect to the mass-scaled atomic positions $\mathbf{u}$, possibly making its evaluation computationally expensive.
This cost becomes particularly significant when the descriptors involve complex functions of atomic coordinates or depend on a large number of atoms, as in membrane fusion.
This introduces a critical bottleneck in training, making the method unfeasible for such systems.
To alleviate this limitation, we adopt the reformulation we recently introduced in Ref.~\citenum{trizio2026ceci}, in which the coordinate gradients present in the original variational functional are restricted to the descriptor space (i.e., $\nabla_\textbf{u} \rightarrow \nabla_\textbf{d}$), thus leading to the approximated and much easier to compute variational functional 
    \begin{equation}
        \label{eq:new_kfunctional}
        \tilde{K}[q(d(\mathbf{x}))] = \left\langle \left| \nabla_{\mathbf{d}} q(d(\mathbf{x})) \right|^4 \right\rangle
    \end{equation}
which can be shown to provide an upper bound to the original functional and is used to define the new variational loss in the training procedure.
Although this new approach does not formally target the \textit{exact} committor function, it has been shown that minimizing the modified functional in Eq.~\ref{eq:new_kfunctional} yields a committor model that remains effective in driving efficient enhanced sampling simulations. 
As a result, this method provides a reliable and computationally efficient alternative, enabling a substantial reduction in training cost for complex systems.

% \subsection{Enhanced sampling scheme}
% \label{sec:enhanced}

\paragraphtitle{Method: Enhanced sampling scheme.}
Effective training of the committor model $q_{\boldsymbol{\theta}}(\mathbf{x})$ requires a dataset containing configurations over the whole relevant phase space, including both the metastable basins and the TS region.
To this end, two complementary bias potentials are applied simultaneously, as proposed in Ref.~\citenum{trizio2025everything}.
The first is the transition-state-oriented Kolmogorov bias introduced in Ref.~\citenum{kang2024computing},
    \begin{equation}
        V_K (\mathbf{x})= -\frac{\lambda}{\beta} \log \left(|\nabla_\textbf{d} q(\mathbf{x})|^2 + \epsilon \right)
    \label{eq:kbias}
    \end{equation}
where $\beta = (k_B T)^{-1}$ is the inverse temperature, $\lambda$ controls the strength of the bias, $\epsilon$ is a regularization parameter, and the gradients $\nabla_\textbf{d} q(\mathbf{x})$ are calculated with respect to the descriptors, as proposed in Ref.~\citenum{kang2024computing}.
The bias $V_K(\mathbf{x})$ is attractive in regions where $|\nabla q(\mathbf{x})|$ is large, i.e., in the TS region, and therefore enhances its sampling.
The $\lambda$ parameter is tuned to yield a bias potential $V_K (\mathbf{x})$ with a magnitude comparable to that of the expected free energy barrier, resulting in efficient sampling of the TS region.

However, $V_K(\mathbf{x})$ alone does not ensure an efficient exploration of the metastable states. 
For this reason, it is combined with the metadynamics-like On-the-fly Probability Enhanced Sampling (OPES) method~\cite{invernizzi2020rethinking, invernizzi2022exploration, trizio2024advanced}, which works by progressively building a repulsive bias that discourages the sampling of already visited configurations.
The OPES bias is defined in the space of a chosen collective variable (CV, $\textbf{s}$), which is a function of the atomic coordinates, i.e.,  $\textbf{s} = \textbf{s}(\textbf{x})$.
In practice, the OPES bias is built so as to drive sampling towards a target distribution $p^{\mathrm{tg}}(\textbf{s})$ in which the probability of observing reactive events is enhanced.
To achieve this, OPES iteratively builds an estimate of the system's probability distribution along the CV and constructs a bias potential that drives the sampled distribution toward $p^{\mathrm{tg}}(\textbf{s})$. 
In practice, we apply the OPES-\textit{explore}~\cite{invernizzi2022exploration} bias $V_{\mathrm{OPES}}(\textbf{x})$ on the committor-related variable $z(\mathbf{x})$ introduced in Ref.~\citenum{trizio2025everything},  which is related to the committor via $q(\mathbf{x}) = \sigma\!\left(z(\mathbf{x})\right)$, where $\sigma$ is a sigmoid-like activation function. 
This change of variable is performed because the sharp behavior of $q(\mathbf{x})$ makes it unsuitable as a CV for enhanced sampling, at variance with $z(\mathbf{x})$, which encodes the same physical information in a smoother form.

The combined action of $V_K(\mathbf{x})$ and $V_{\mathrm{OPES}}(\textbf{x})$ results in a sampling scheme that uniformly covers the full reactive process, simultaneously promoting transitions between metastable states and increasing sampling of the crucial transition-state region.

% \subsection{Self-consistent iterative procedure}
% \label{sec:iterative}

\paragraphtitle{Method: Self-consistent iterative procedure.}
The neural network parametrization and the enhanced sampling scheme are combined into the self-consistent iterative procedure summarized here~\cite{kang2024computing,trizio2025everything}.

\begin{itemize}
\item \textbf{Step 1.} At iteration $n$, the committor model $q_{\boldsymbol{\theta}}^n(\mathbf{x})$ is trained on the dataset of configurations $\mathbf{x}^n$ and weights $\textbf{w}^n$ available at that point.
At the first iteration ($n=0$), the model is trained only on the labeled 
configurations, obtained by running short unbiased simulations in the metastable basins.

\item \textbf{Step 2.} Enhanced sampling simulations are performed under the combined action of $V_\mathrm{OPES}^n (\textbf{x})$, applied on the committor-related CV $z(\textbf{x})$, and the Kolmogorov bias $V_K^n (\textbf{x})$.

\item \textbf{Step 3.} The configurations sampled at Step 2 are treated as the new training set, after being reweighted according to the effective bias $V_{\text{eff}}^n (\textbf{x}) = V_{K}^n (\textbf{x}) + V_{\text{OPES}}^n (\textbf{x})$ with weights given by $w_i^{n} =e^{\beta\,V_{\mathrm{eff}}^{n}(\textbf{x}_{i})} / {\left\langle e^{\beta\,V_{\mathrm{eff}}^{n}(\textbf{x})}\right\rangle_{U_{\mathrm{eff}}^{n}}}$. 

\end{itemize}
The above steps are repeated iteratively to progressively refine both the committor estimate and the sampling, until convergence is achieved.\\

\paragraphtitle{Results.}
As anticipated, we investigate stalk formation in water between two parallel (25 $\times$ 25) nm membrane patches, both composed of 1,2-dioleoyl-\textit{sn-glycero}-3-phosphocholine (DOPC) lipids. 
The process is mediated by an Au nanoparticle (AuNP) with a core diameter of 2 nm, functionalized with 11-mercapto-1-undecanesulfonate (MUS) and octanethiol (OT) ligands, and embedded in the lower bilayer. 
The functionalized AuNP, the lipids and the solvent were modeled at coarse-grained level with the Martini 2~\cite{marrink2007MARTINI} force field, as detailed in our previous works~\cite{simonelli2015monolayer, brosio2023nanoparticle}.
The system was hydrated with a water-to-lipid ratio of about 12 Martini water beads per lipid molecule.

Preliminary simulations reveal that, at the level of dehydration considered here, before the stalk is formed the membranes can adopt two distinct configurations. In the pre-stalk state (state $A$), the membrane is locally curved and the NP ligands remain in contact with the upper bilayer, maintaining the two membranes in close apposition. In a secondary state $A'$, instead, the ligands detach from the upper leaflet, and the membranes relax toward a flatter configuration.
Working under moderate, rather than extreme, dehydration is desirable because it allows the membranes to retain more natural shape and thickness fluctuations, and is therefore expected to provide a more realistic description of the local rearrangements leading to stalk formation. At the same time, these fluctuations make it possible for the system to leave the pre-stalk state by membrane detachment. Since our aim is to sample the transition from the pre-stalk state $A$ to the stalk state $B$, rather than the reversible association and dissociation of the two membranes $A \leftrightarrow A'$, we prevent visits to the detached $A'$ state by applying a lower-wall restraint on the number of contacts between the terminal ligand beads and the lipid headgroups of the upper membrane.

Because NP-mediated stalk formation is localized in the intermembrane region above the embedded NP, we designed a set of input descriptors for the committor model to account for the presence of hydrophobic tails in this region. 
As stalk formation is initiated by the accumulation and reorientation of lipid tails between the two bilayers, we define the descriptors as the number of lipid tail beads contained in 17 concentric spherical shells centered on the NP and restricted to the inter-membrane region (Fig.~\ref{fig:descriptors}).
This can be seen as a generalization of the geometric construction of the chain coordinate \cite{awasthi2016simulations, hub2017probinga, hub2021joint}, without the imposition of a pre-defined directionality to the orientation of the stalk structure.
In addition, we apply two geometric restrictions to the shells to consider in the calculations only the tails actively contributing to stalk formation. 
On one hand, a height cutoff that excludes lipids below the surface of the NP, and on the other, a lateral cutoff of 2~nm from the NP center, to avoid the application of bias to regions too far from the NP. 
Together, this localizes the descriptors onto the region where the hydrophobic bridge forms. Conveniently, this overall construction also allows using a single shared neighbor list, which minimizes the computational overhead associated with descriptor evaluation. 

Once the descriptor set was chosen, we trained a committor model $q_{\boldsymbol{\theta}}$ using the self-consistent iterative protocol of Ref.~\citenum{trizio2026ceci}, which we fully detail in the Methods. A test of the robustness of the committor against alternative choices of the descriptors in terms of number and thickness of shells is reported in Fig. \ref{SI_fig:descriptor_comparison}.

\begin{figure}[h!]
    \centering
    \includegraphics[width=1\linewidth]{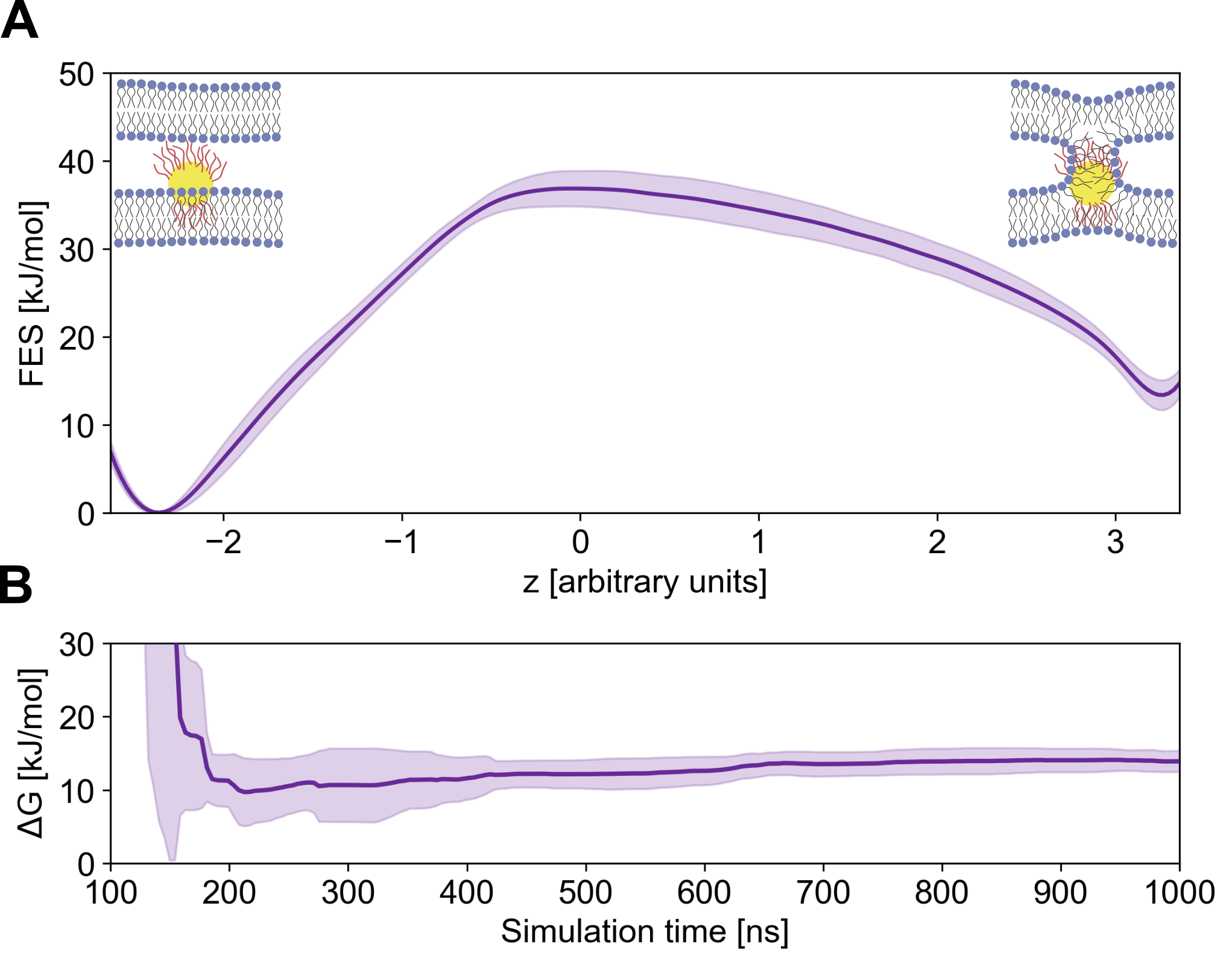}
    \caption{\textbf{Free energy estimates.} \textbf{A}: Free energy surface (FES) for the stalk formation projected on the committor-related CV $z(\mathbf{x})$; the insets schematically depict the two states and are placed above the corresponding CV values. \textbf{B}: Estimate of the free energy difference $\Delta G$ between the two states as a function of simulation time.
     In both panels, the solid line reports the average over 5 independent 1 $\mu s$ replicas, while shaded regions indicate the associated statistical errors. 
     In both cases, an initial transient of 100 $ns$ has been discarded.} 
    \label{fig:fes}
\end{figure}

\begin{figure*}[t]
    \centering\includegraphics[width=1\linewidth]{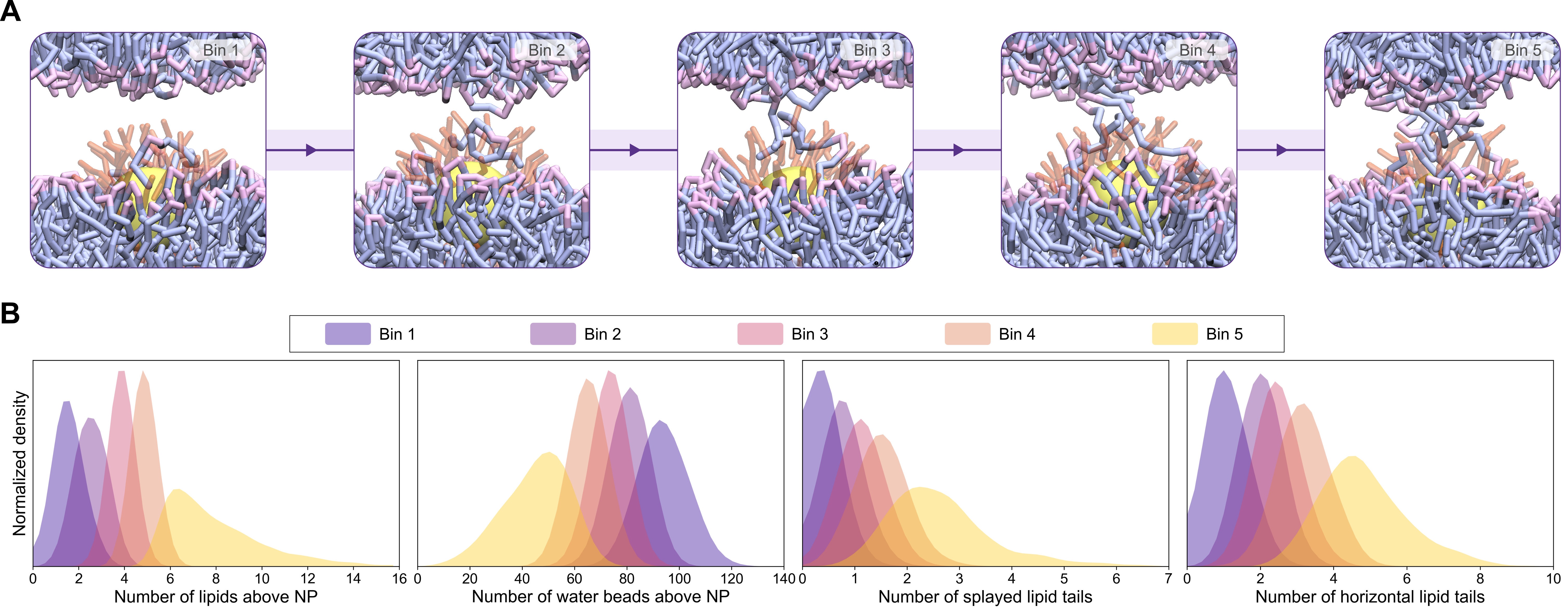}
    \caption{\textbf{A}: Representative medoid configurations from five bins ordered according to their committor value, showing the progressive formation of the stalk between the two bilayers. \textbf{B}: Normalized distributions of four structural descriptors across the five committor bins, respectively: number of lipids above the NP, number of water beads above the NP, number of splayed lipid tails, and number of horizontally oriented tail segments.}\label{fig:bindistributions}
\end{figure*}

Enhanced sampling is driven by two complementary bias potentials.
On one hand, the Kolmogorov bias $V_K(\mathbf{x})$ enhances the sampling of the transition-state ensemble, on the other, the \textit{explore} variant of the metadynamics-like OPES bias~\cite{invernizzi2020rethinking,invernizzi2022exploration} progressively fills the free energy landscape along the committor-related CV $z(\mathbf{x})$, promoting transitions between the pre-stalk and stalk basins. 

At convergence, which was reached after two iterations, we used the committor model thus obtained to perform a set of 5 independent 1 $\mu s$ production simulations, to gather meaningful statistics.
The free energy estimates obtained via reweighting from these calculations are reported in Fig.~\ref{fig:fes}. 
These are compatible with results obtained using the chain coordinate~\cite{brosio2023nanoparticle} (see SI) and show a consistent behavior across all replicas, thus confirming the reliability of our calculations.

Beyond free-energy reconstruction, the learned committor provides insight into the reaction mechanism, since the variable $z(\textbf{x})$ naturally orders the reaction progress~\cite{kang2024computing,trizio2025everything}. 
We thus grouped the sampled configurations into five bins according to their $z(\textbf{x})$ value, ordered from the no-stalk basin to the stalk basin.
For each bin, we identified the most representative structure using k-medoids clustering~\cite{Schubert2022kmedoids} in a space combining the lipid-tail descriptors with an equivalent set of water-coordination variables, defined using the same shell geometry (Fig.~\ref{fig:bindistributions}A). 
Inspection of these representative structures reveals the gradual rearrangement of the lipids along the transition. In the transition-state bin, the hydrophobic connection involves only a few lipids: one lipid from the lower membrane protrudes upward, while a lipid from the upper membrane exposes one tail toward the intermembrane region, together with a partially detached lipid located between the two leaflets.
In the following bin, the connection broadens through the recruitment of lipids from both membranes, before evolving into the more compact and continuous hydrophobic structure observed in the stalk basin.
The progression also highlights the different roles of the two membranes. The lower membrane undergoes strong local fluctuations, with repeated lipid-tail protrusions toward the intermembrane space as a consequence of the NP-induced perturbation. 
These fluctuations alone, however, are not sufficient to initiate stalk formation. 
The transition progresses only when a lipid from the upper membrane reorients one of its tails toward this region, establishing the first hydrophobic contact between the two leaflets (see Fig.~\ref{SI_fig:contacts}).
It is worth noting that this analysis is made possible by the uniform OPES+$V_K(\mathbf{x})$ sampling, which provides a large number of configurations across the whole transition pathway, including the TS region, retaining sufficient statistics in each bin.

Beyond the qualitative information that single structures provide, we also characterize the process quantitatively, by computing ensemble averages of several quantities relevant to stalk formation dynamics, which report on the local hydration and lipid rearrangement above the NP (Fig.~\ref{fig:bindistributions}B).
Moving from the pre-stalk basin to the stalk basin, the number of lipid tails above the NP increases, while the number of water beads above the NP decreases. 
In addition, the transition is accompanied by the presence of an increasing number of splayed lipid tails and horizontally oriented tails, consistent with the collective lipid reorientation and protrusion events that precede hydrophobic stalk formation. 
Together, these trends show how stalk formation proceeds through the combined expulsion of water, recruitment of lipid tails from both leaflets, and their progressive reorientation into a continuous hydrophobic connection, underlying the collective nature of the transition.

% \section{Discussion}
\paragraphtitle{Discussion.}
In this work, we applied our recent committor-based enhanced sampling framework to study NP-mediated stalk formation. 
Starting from information limited to the pre-stalk and stalk states only, the committor estimate was progressively refined through alternating cycles of learning and enhanced sampling, and the resulting committor-related variable was used to promote repeated transitions across the full pathway, while enhancing the sampling of the transition-state region. 
This resulted in a uniform exploration of the reactive process, converged free-energy estimates, and mechanistic insight into the transition mechanism. 

As with any other machine-learning-based method, the effectiveness of our approach depends on the choice of the descriptors used as inputs to the model. 
In this regard, the concentric lipid-tail coordination shells adopted here, which preserve the physical information of hydrophobic-connectivity CVs, represent a flexible option that can be applied to describe fusion processes mediated not only by NPs but also by other fusogenic peptides, proteins, and synthetic molecules.
In addition, our approach is also robust and not too sensitive with respect to the choice of the shell geometry, as proved by the agreement between models trained with different shell numbers and thicknesses, and the descriptors are also computationally efficient, as all shells share a common neighbor list.

Even if for the studied system these descriptors proved informative enough to obtain good sampling and converged free-energy estimates, thus representing a promising option also for more complex systems, we note that in such cases additional descriptors could also be included to better capture finer and distinct degrees of freedom to further accelerate the dynamics.
For example, descriptors related to membrane curvature, lipid orientation, or specific fusogen-lipid interactions could be incorporated to study mechanisms in which the transition mechanism is more strongly related to membrane deformation or molecular recognition.
It is also worth noting that the vertical and lateral cutoffs used here deliberately focus the descriptors on the region above the NP, where stalk formation is expected to occur. 
While this introduces a reasonable and somewhat general prior information about the location and direction of the stalk during the process, these restrictions could be relaxed and adapted to different fusogen geometries to provide a less directional representation in which the position and orientation of the hydrophobic connection emerge directly from the simulations.

Beyond its sampling efficiency and flexibility, the probabilistic description provided by the committor is particularly well suited to describe the complexity of membrane fusion, in which the reaction progress emerges from several coupled and collective molecular rearrangements rather than from a single event. 
Ordering configurations along the committor shows that the transition proceeds when lipid tails from the upper membrane reorient toward the hydrophobic region above the NP and establish the first hydrophobic contact. 
This is accompanied by gradual water depletion and by the recruitment, splaying, and reorientation of multiple lipids, supporting a cooperative mechanism of stalk formation.

Overall, this work shows a general, semi-automatic and efficient framework for the study of stalk formation, which, starting from very limited information, leads to deep insights into fusion mechanisms and thermodynamics.
We believe that this, combined with the flexibility of the whole procedure, will provide a precious tool to better understand fusion processes in realistic conditions.

%% file: manuscript/supporting.tex
\section{Stalk formation}
    \subsection{Computational details}
        \paragraphtitle{Simulation details}
        All simulations were carried out targeting the NPT ensemble, using the GROMACS-2024.5~\cite{abraham2015gromacs} MD engine patched with PLUMED-2.9.4~\cite{tribello2014plumed, plumed2019promoting} and the Martini 2~\cite{marrink2007MARTINI} force field. 
        
        Two apposed DOPC bilayers separated by an initial distance of 3 nm, in a (25 $\times$ 25 $\times$ 15) nm simulation box were simulated in the isothermal–isobaric (NPT) ensemble using a time step of 20 fs. Temperature was maintained at 310 K using the velocity-rescale thermostat~\cite{bussi2007velocity} with a coupling time constant of 1.0 ps. 
        Pressure was set at 1 bar using the Parrinello–Rahman barostat with semi-isotropic pressure coupling, and a coupling time constant of 12 ps. 
        Periodic boundary conditions were applied in all three spatial directions.

        \paragraphtitle{Committor model training details}       
        To model the approximated committor, we used a set of 17 coordination numbers as inputs of a neural network (NN) with architecture [17, 14, 12, 1] nodes/layer and an initial normalization layer. 
        Starting from the standard \texttt{COORDINATION} action in PLUMED, we implemented a custom multi-radius coordination function, \texttt{COORDINATION\_MULTI}. This action computes smooth coordination numbers between the NP center of mass and the lipid tail beads for multiple radii in a single call.
        \begin{lstlisting}[language=bash,caption={Example of PLUMED input file for \texttt{COORDINATION\_MULTI} action}, captionpos=b,label={lst:coordination}]
        # Define the nanoparticle group from the index file
        NP: GROUP NDX_FILE=index.ndx NDX_GROUP=NP
        
        # Define the lipid tail bead group from the index file
        tails: GROUP NDX_FILE=index.ndx NDX_GROUP=tails
        
        # Compute the nanoparticle center of mass
        com: CENTER ATOMS=NP
        
        # Compute multi-radius coordination numbers between the NP COM and lipid tails
        c_tails: COORDINATION_MULTI GROUPA=com GROUPB=tails D_0=0.0
        R_0=1.6,1.7,1.8,1.9,2.0,2.1,2.2,2.3,2.4,2.5,2.6,2.7,2.8,2.9,3.0,3.1,3.2,3.3
        NN=50 MM=100 D_MAX=3.6 NLIST NL_CUTOFF=3.8 NL_STRIDE=10 N_OUT=18 Z_SHIFT=1 XY_RADIUS=2
        \end{lstlisting}
        
        Here, \texttt{GROUPA=com} defines the NP center of mass and \texttt{GROUPB=tails} the target lipid tail beads, the values in \texttt{R\_0} define concentric spheres radii around the NP, and \texttt{N\_OUT=18} sets the number of output coordination numbers, one per shell.
        For efficiency, the neighbor list is built using the largest cutoff radius and then reused for all smaller radii. 
        Shell-specific contributions are then obtained separately in the PLUMED input file by subtracting the coordination values computed for consecutive radii.
        The calculation is restricted to the region above the NP by \texttt{Z\_SHIFT=1}, which excludes beads below a plane 1 nm above the NP COM, and by \texttt{XY\_RADIUS=2}, which retains only beads within 2 nm laterally from the NP center. 
        The complete implementation is provided in the code availability section.
        All keywords not explicitly discussed here correspond to the standard parameters of the PLUMED \texttt{COORDINATION} action.\\
        
        OPES+$V_K$ simulations biasing the committor-based $z$ variable, with OPES Explore \texttt{PACE} = 500, were performed. 
         During all production simulations, a lower wall was applied to the coordination number between the terminal ligand beads, \texttt{MUS4}, and the lipid head beads. 
        This restraint was used to maintain a minimum number of ligand-upper membrane contacts and prevent the system from entering a third substate in which the membranes flatten, and the ligands lose contact with the upper membrane. \\
        
        For the committor training, from the first iteration on, we discarded the first part of the simulations in which the bias was stabilizing.
        For the optimization, we used the ADAM optimizer with an initial learning rate of $10^{-3}$ modulated by an exponential decay with multiplicative factor $\gamma=0.99995$.
        The training was performed for 10000 epochs during the first iteration, for 5000 during the second one, and for 40000 epochs during the final one. 
        The $\alpha$ hyperparameter in the loss function was set to 10 for the first iteration and to 10$^-3$ for the other two, while the $\gamma$ hyperparameter was set to 1 for the first iteration and to $10^{3}$ for the other two iterations.  We optimized the $\log$ of the total loss for numerical stability. 
         For each iteration, Table~\ref{sup_tab:iteration} reports the size of the training dataset, the simulation time $t_d$ from which this dataset was collected, the OPES Explore \texttt{BARRIER}, the value of $\lambda$, and the production simulation time $t_s$ performed with the trained model.
        In this case, only data from the previous iteration was used for the training.

        \begin {table}[h!]
                \caption {Summary of the iterative procedure.} \label{sup_tab:iteration}
                %\vspace{-5mm}
                \begin{center}
                \begin{tabular}{ |c|c|c|c|c|c|c| } 
                 \hline
                 Iteration & Dataset size & $t_d$ [ns] & OPES \texttt{BARRIER} [kJ/mol] & $\lambda$ & $t_s$ [ns] \\ 
                 \hline
                    0   & 10000 & 2*1000 & 30& 30  & 5*500\\
                    1   & 59000 & 5*500 & 30& 30  & 5*500 \\
                    2   & 97500 & 5*500 & 30& 50  & 5*1000 \\
                 \hline
                \end{tabular}
                \end{center}
            \end {table}
        
    \subsection{Additional figures}
    
        \begin{figure}[h!]
            \centering
            \includegraphics[width=0.8\linewidth]{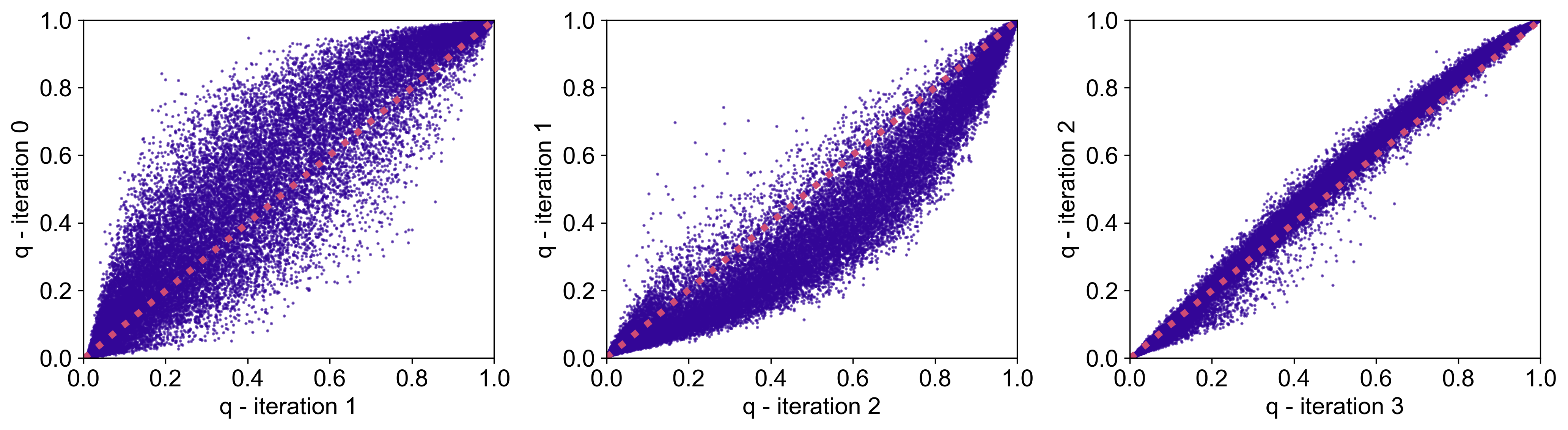}
            \caption{\textbf{Convergence of the committor model} across consecutive training iterations. Each panel compares the committor values predicted by two successive iterations, with the dashed red line indicating perfect agreement. The scatter progressively collapses around the diagonal, showing that the model converges over successive iterations. The committor obtained at iteration 2 was used to drive production simulations, while the model was subsequently retrained to verify convergence and to assign configurations to committor-based bins.
            } 
            \label{SI_fig:iter_comparison}
        \end{figure}
        
        \begin{figure}[h!]
            \centering
            \includegraphics[width=1\linewidth]{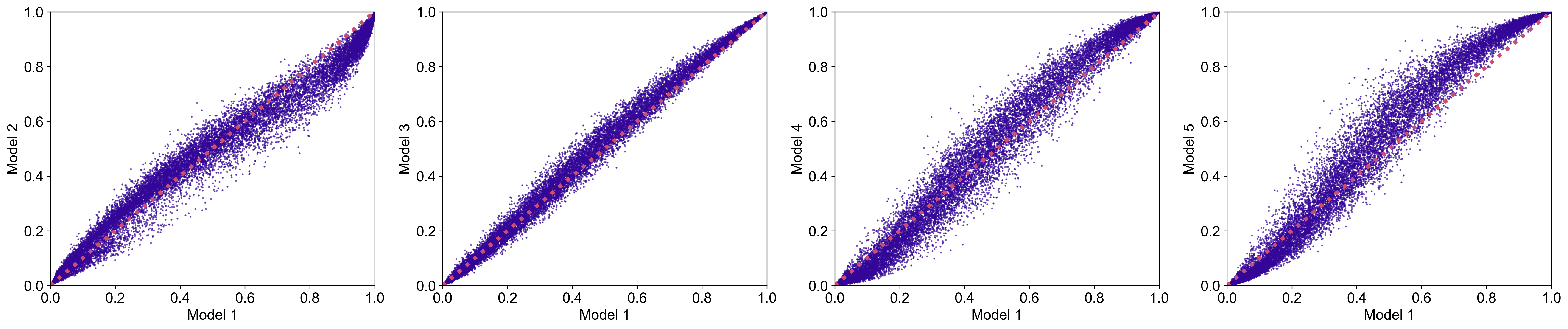}
            \caption{\textbf{Agreement between committor models trained with different descriptor sets.} Model 1 is the reference model used in production simulations and was trained with the selected descriptor setup of 17 lipid tail coordination shells with a width of 0.1 nm. It is compared against alternative models trained with 34 shells of 0.05 nm (model 2), 12 shells of 0.15 nm (model 3), 8 shells of 0.2 nm (model 4), and 6 shells of 0.3 nm (model 5), respectively. The $x$-axis reports the committor predicted by model 1, while the $y$-axis reports the committor predicted by each alternative model. The dashed red line represents perfect agreement. The overall alignment of the points with the diagonal indicates that the committor prediction is not overly sensitive to reasonable changes in the descriptor sets.
            } 
            \label{SI_fig:descriptor_comparison}
        \end{figure}

        \begin{figure}[h!]
            \centering
            \includegraphics[width=0.45\linewidth]{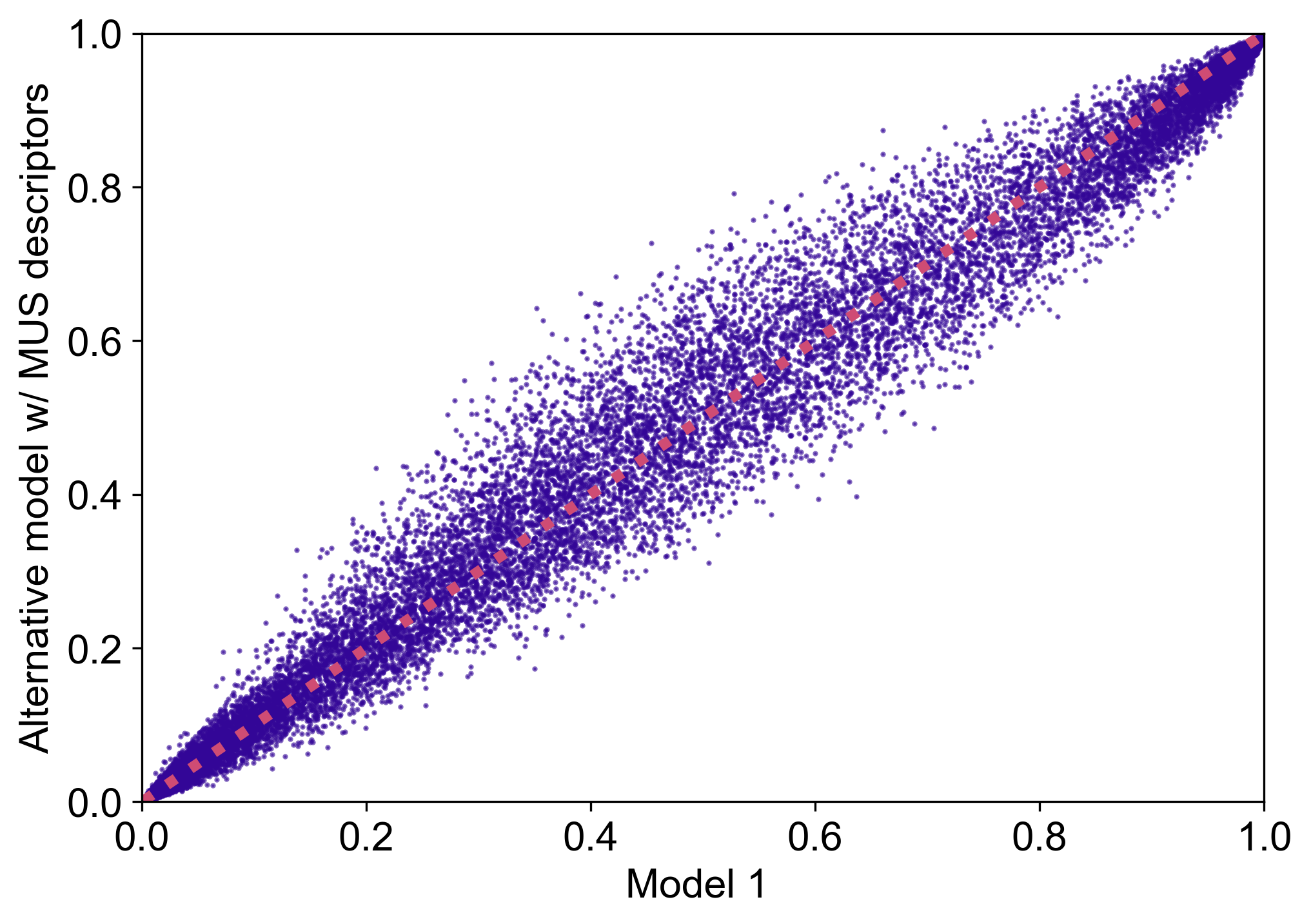}
            \caption{\textbf{Agreement between committor models trained with different descriptor sets.} Model 1 is the reference model used in production simulations and was trained with the selected descriptor setup of 17 lipid tail coordination shells with a width of 0.1 nm. 
            This model is compared with an alternative model trained using the same 17 lipid-tail descriptors, supplemented by 6 MUS ligand coordination shells with a width of 0.2 nm, and a lateral cutoff of 1nm with respect to the NP core.
            The $x$-axis reports the committor predicted by model 1, while the $y$-axis reports the committor predicted by the alternative model. 
            The dashed red line represents perfect agreement. The overall alignment of the points with the diagonal indicates that the committor prediction is not overly sensitive to reasonable changes in the descriptor sets.
            } 
            \label{SI_fig:descriptor_comparison_mus}
        \end{figure}

\newpage

    \subsection{Chain coordinate}
        In this work, we used a re-adapted version of the chain coordinate originally introduced by Hub and Awasthi~\cite{hub2017probinga} and implemented in PLUMED by Di Bartolo and Masone~\cite{bartolo2022synaptotagmin1}. 
        Here, the CV was modified to describe NP-mediated stalk formation through the custom PLUMED action \texttt{MEMFUSIONP\_NP}, available in the code repository.
        The action was used in the simulations with the following PLUMED input.
        \begin{lstlisting}[language=bash,caption={Example of PLUMED input file for \texttt{MEMFUSIONP\_NP} action}, captionpos=b,label={lst:memfusion}]
        # Define the nanoparticle group from the index file
        NP: GROUP NDX_FILE=index.ndx NDX_GROUP=NP
        
        # Define the lipid tail bead group from the index file
        tails: GROUP NDX_FILE=index.ndx NDX_GROUP=tails
        
        # Compute the nanoparticle center of mass
        com: CENTER ATOMS=NP

         # Compute the chain coordinate
         memFusion: MEMFUSIONP_NP NP=com TAILS=tails NSMEM=20 DSMEM=0.1 ZSHIFT=1.1 RCYLMEM=2.0
        \end{lstlisting}
        Here, \texttt{NP=com} defines the NP center of mass, while \texttt{TAILS=tails} specifies the lipid tail beads used to evaluate the coordinate. 
        The parameters \texttt{NSMEM=20} and \texttt{DSMEM=0.1} define 20 slices separated by 0.1 nm along the membrane normal, corresponding to a total sampling height of 2.0 nm. 
        Compared with the original implementation, the cylindrical region is anchored to the NP center of mass instead of being defined from membrane reference groups. 
        Its vertical position is controlled by the new parameter \texttt{ZSHIFT=1.1}, which places the cylinder above the NP, while \texttt{RCYLMEM=2.0} restricts the calculation laterally to a cylindrical region of radius 2.0 nm around the NP center. 
        
        We performed 5 $\times$ 1 $\mu s$ OPES Explore replicas using this adapted chain coordinate as CV, with a \texttt{BARRIER} parameter of 30 kJ mol$^{-1}$.
     
        \begin{figure}[h!]
            \centering
            \includegraphics[width=0.6\linewidth]{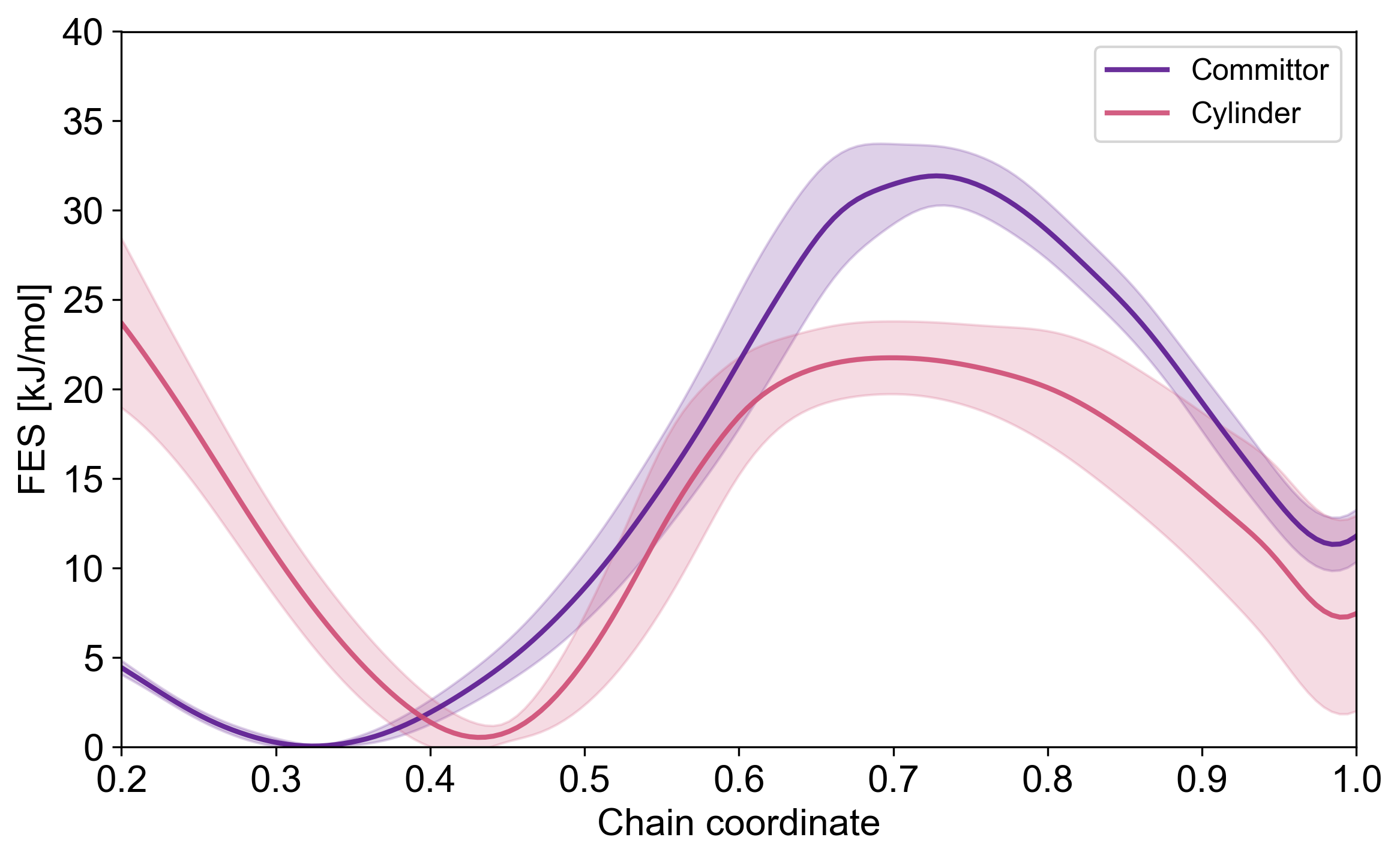}
            \caption{\textbf{Comparison between committor and chain-coordinate free energy estimates.} Free energy surface (FES) for stalk formation. For comparison they are both projected on the chain coordinate CV. The committor-based simulations are shown in purple, while the chain-coordinate simulations are shown in pink. Solid lines report the average over five independent 1$\mu$s replicas, and shaded regions indicate the associated statistical errors. An initial transient of 100 ns was discarded.
            } 
            \label{SI_fig:fes_comparison}
        \end{figure}

        \begin{figure}[h!]
            \centering
            \includegraphics[width=0.7\linewidth]{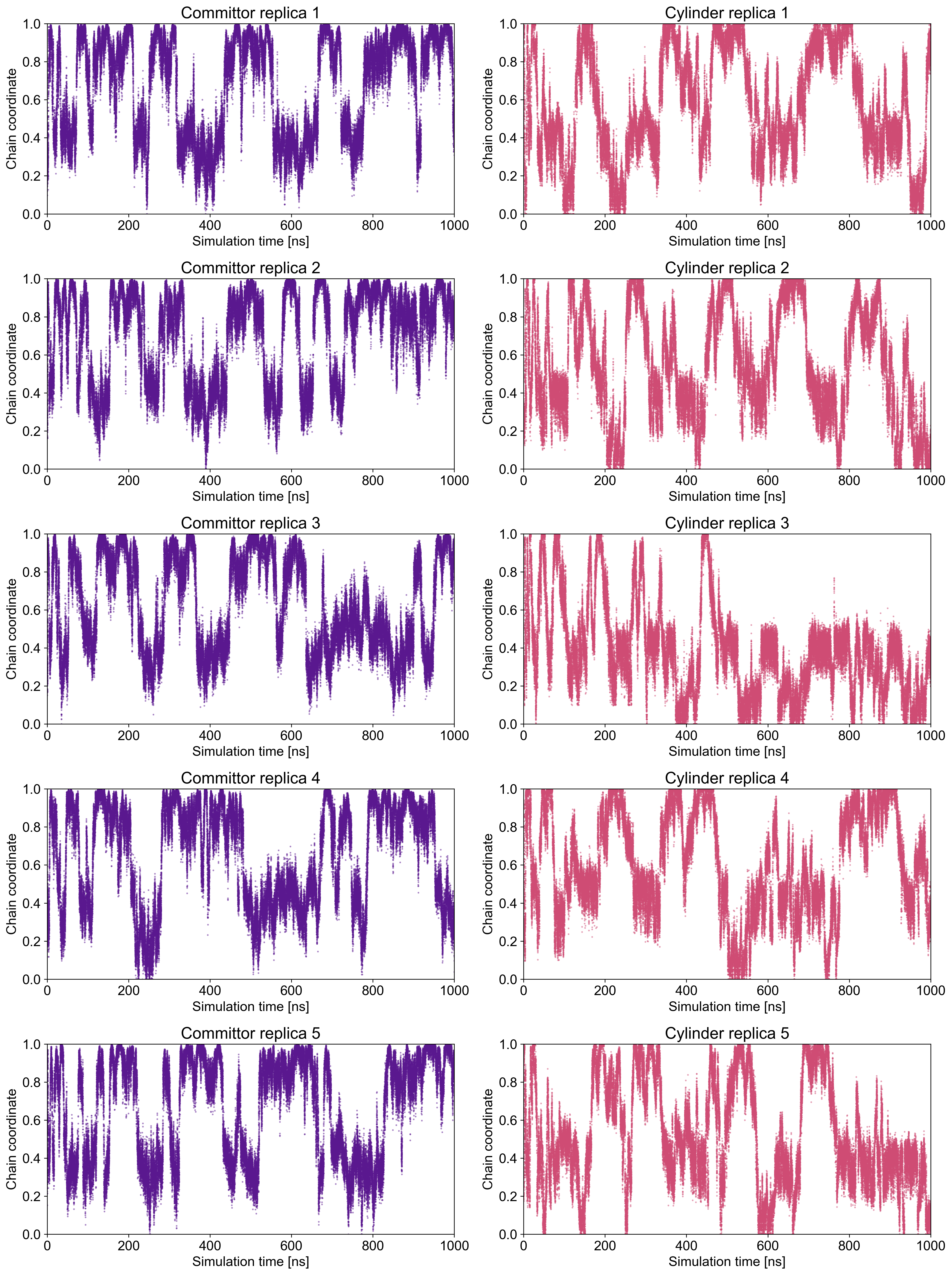}
            \caption{\textbf{Comparison of sampling obtained with committor-based and chain-coordinate approaches.} Time series of the chain coordinate for five independent 1 $\mu$s simulations performed using the committor-based framework (left, purple) and using the chain coordinate (right, pink). For reference, no-stalk configurations correspond to chain coordinate values $<0.5$, whereas stalk states correspond to values $>0.8$.} 
            \label{SI_fig:transitions}
        \end{figure}

         \begin{figure}[h!]
            \centering
            \includegraphics[width=1\linewidth]{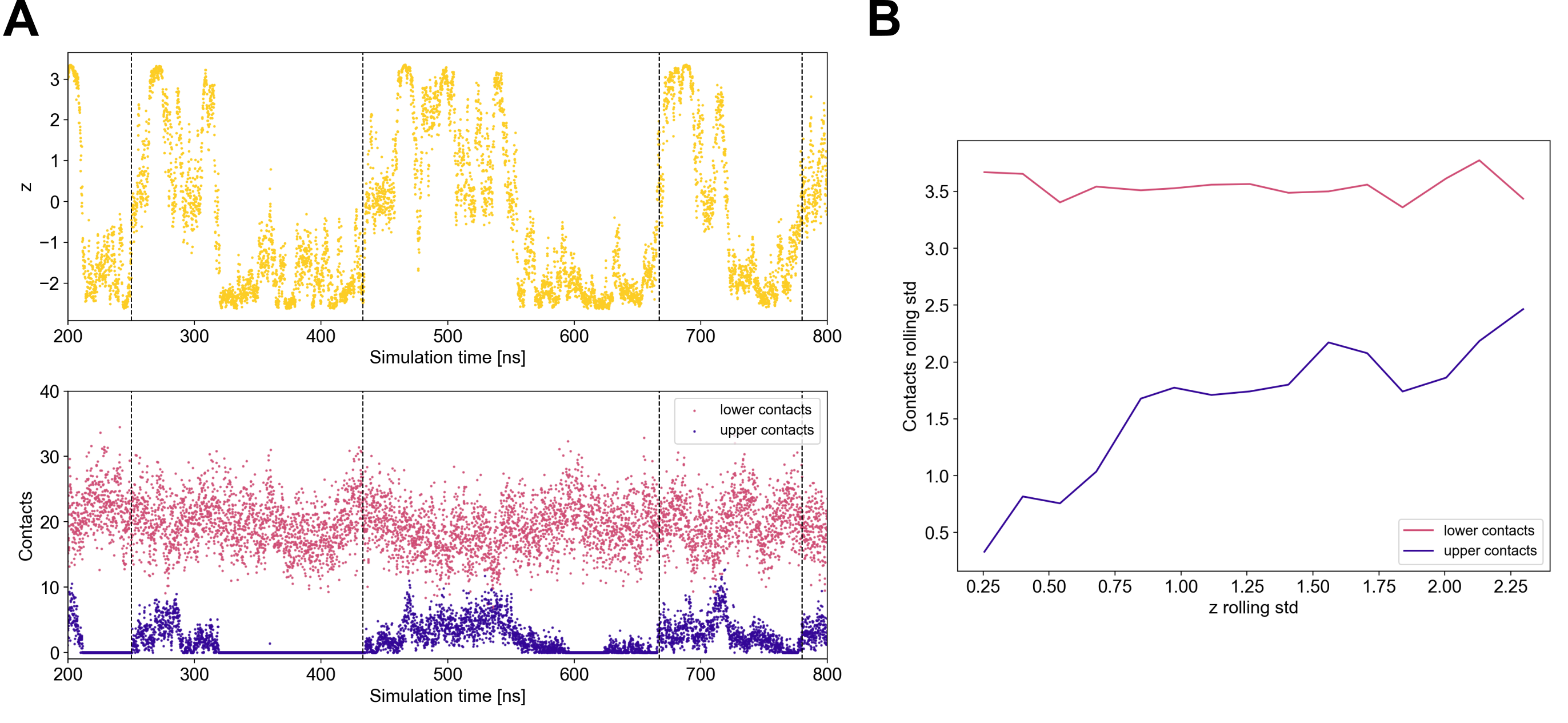}
            \caption{\textbf{Correlation between hydrophobic tail–ligand contacts and fluctuations of the committor variable.} \textbf{A:} Time evolution of the committor-related variable z (top, yellow) and of the number of contacts between lipid terminal tail beads and the hydrophobic beads of the NP ligands (bottom). Contacts are reported separately for lipids belonging to the lower membrane (pink) and upper membrane (purple). \textbf{B:} Rolling standard deviation of the lower- and upper-membrane contacts as a function of the rolling standard deviation of z. Fluctuations in z are accompanied by increasing fluctuations in upper-membrane contacts, whereas lower-membrane contacts remain constant. This suggests that rearrangements of the upper membrane are more closely associated with progress along the transition, while the lower membrane remains continuously perturbed by the embedded NP.} 
            \label{SI_fig:contacts}
        \end{figure}